\documentclass[a4paper]{article}

\usepackage[utf8]{inputenc}
\usepackage[english]{babel}
\usepackage{amsmath,amssymb,mathtools, bm}
\usepackage{pifont} % http://ctan.org/pkg/pifont
\usepackage{xspace}
\usepackage{geometry}
\usepackage{placeins}
\usepackage{booktabs,multirow,makecell,array,tabularx}
\usepackage[font=small,labelfont=bf]{caption}
\usepackage[subrefformat=parens,labelformat=parens]{subcaption}
\usepackage[dvipsnames]{xcolor}
\usepackage{tikz,pgfplots,pgfplotstable}
\usetikzlibrary{automata,er,patterns,calc,external}
\usepgfplotslibrary{fillbetween}
\usepackage[mode=buildnew]{standalone}
\usepackage{siunitx}
\usepackage{csquotes}
\usepackage[nolist]{acronym}
\usepackage[
bibstyle=cce,giveninits=true,date=year,articlein=false,innamebeforetitle=true,isbn=false,sorting=nyt,
  citestyle=authoryear-comp,uniquelist=minyear,uniquename=init,sortcites=false,
  backend=biber
 ]{biblatex}
\usepackage{hyperref}
\usepackage[capitalize]{cleveref}
\makeatletter
\newcommand{\customlabel}[2]{%
\protected@write \@auxout {}{\string \newlabel {#1}{{#2}{}}}}
\makeatother
\tikzstyle{insetplotmark}=[thick,densely dotted,pattern=north west lines, pattern color=black,opacity=.50]
\pgfplotsset{
	plotbenchdataaxis/.style={
		grid=major, xlabel={step}, ylabel={goal function},legend style={at={(0.5,1.025),font=\small},anchor=south}, legend cell align=left,
	},
	plotbenchtimeaxis/.style={
		grid=major, xlabel={step}, ylabel={time [s]},legend style={at={(0.5,1.025),font=\small},anchor=south}, legend cell align=left,
	}
}

\definecolor{color5A}{HTML}{22046e}
\definecolor{colorL1}{HTML}{085c2b}

\hypersetup{
    colorlinks,
    linkcolor={blue},
    citecolor={black},
    urlcolor={blue}
}
\pgfplotsset{compat=newest}
\crefname{equation}{equation}{equations}    
\providecommand{\keywords}[1]{\small \textbf{Keywords---} #1}
\newcommand{\ie}{i.\,e.\@\xspace}
\newcommand{\eg}{e.\,g.\@\xspace}
\newcommand{\cf}{cf.\@\xspace}

\newcommand{\wrt}{w.r.t.\@\xspace}
\newcommand{\xmark}{\ding{55}}%

\begin{acronym}	
    \acro{TRSP}[TRSP]{\emph{transport robot scheduling problem}}
    \acro{QC}[QC]{quantum computing}
    \acro{QUBO}[QUBO]{quadratic unconstrained binary optimization}
    \acro{GPGPU}[GPGPU]{general purpose computation on graphics processing unit}
    \acro{MIP}[MIP]{mixed integer program}
    \acro{LP}[LP]{linear programming}
    \acro{AGV}[AGV]{automated guided vehicles}
    \acro{JSSP}[JSSP]{job shop scheduling problem} 
    \acro{QAOA}[QAOA]{quantum approximate optimization algorithm}
    \acro{MIS}[MIS]{maximum independent set}
    \acro{BQM}[BQM]{binary quadratic model}
    \acro{CQM}[CQM]{constrained quadratic model}
    \acro{LBQM}[LBQM]{D-Wave's hybrid \emph{Leap} framework}
    \acro{FDA}[FDA]{Fujitsu's digital annealer}
    \acro{FDAh}[FDAh]{Fujitsu's digital annealer hybrid framework}
    \acro{TSB}[TSB]{Toshiba's Simulated Bifurcation Machine}
    \acro{API}[API]{application programming interface}
    \acro{NISQ}[NISQ]{noisy intermediate-scale quantum}
    \acro{SG}[SE-GU]{Gurobi with the sequence model}
    \acro{TG}[TI-GU]{Gurobi with the time-indexed model}
    \acro{QLBQM}[QU-LBQM]{\ac{LBQM} with the \acs{QUBO} model}
    \acro{QFDA}[QU-FDA]{\ac{FDA} with the \acs{QUBO} model}
    \acro{QFDAh}[QU-FDAh]{\ac{FDAh} with the \acs{QUBO} model}
    \acro{TG-QLBQM}[\acs{TG}@\acs{QLBQM}]{the relative runtime of \acs{TG} \wrt \acs{QLBQM}}
    \acro{SG-QLBQM}[\acs{SG}@\acs{QLBQM}]{the relative runtime of \acs{SG} \wrt \acs{QLBQM}}
    \acro{TG-QFDA}[\acs{TG}@\acs{QFDA}]{the relative runtime of \acs{TG} \wrt \acs{QFDA}}
    \acro{SG-QFDA}[\acs{SG}@\acs{QFDA}]{the relative runtime of \acs{SG} \wrt \acs{QFDA}}
    \acro{SG-QFDAh}[\acs{SG}@\acs{QFDAh}]{the relative runtime of \acs{SG} \wrt \acs{QFDAh}}
\end{acronym}

\usepackage{todonotes}
\definecolor{notecolorbg}{HTML}{ffe0b3}
\definecolor{notecolorfg}{HTML}{b30000}
\setuptodonotes{inline,backgroundcolor=notecolorbg,textcolor=notecolorfg,size=\small}

\title{An Optimization Case Study for solving a Transport Robot Scheduling Problem on Quantum-Hybrid and Quantum-Inspired Hardware}
\hypersetup{pdftitle={An Optimization Case Study for solving a Transport Robot Scheduling Problem on Quantum-Hybrid and Quantum-Inspired Hardware}}

\author{Dominik Leib\textsuperscript{1}\thanks{D. Leib: dominik.leib@itwm.fraunhofer.de}, Tobias Seidel\textsuperscript{1}, Sven J\"ager\textsuperscript{1}, Raoul Heese\textsuperscript{1}, Caitlin Jones\textsuperscript{2}, \\ Abhishek Awasthi\textsuperscript{2}\thanks{A. Awasthi: abhishek.awasthi@basf.com}, Astrid Niederle\textsuperscript{3}, Michael Bortz\textsuperscript{1}}
\date{
	\textsuperscript{1} Fraunhofer ITWM, Kaiserslautern, Germany\\[1ex]
	\textsuperscript{2} BASF Digital Solutions GmbH, Ludwigshafen am Rhein, Germany\\[1ex]
    \textsuperscript{3} BASF SE, Ludwigshafen am Rhein, Germany\\[1ex]
}

\begin{document}

%Labels for Supplementary Material
\customlabel{sec:models}{Section S1}
\customlabel{subsec:QuboModel}{Section S1.1}
\customlabel{eq:QuboTimeHorizon}{Eq. (S2)}
\customlabel{subsec:SequenceModel}{Section S1.2}
\customlabel{fig:routestimeindexed}{Fig. S1}
\customlabel{subsec:TimeIndexedModel}{Section S1.3}
\customlabel{sec:benchmark library}{Section S2}
\customlabel{fig:NumVariablesPerBox}
{Fig. S2}
\customlabel{sec:solver settings}{Section S3}
\customlabel{sec:prestudy}{Section S4}
\customlabel{eq:QuboModel}{Eq. (S16)}

\maketitle

\begin{abstract}
We present a comprehensive case study comparing the performance of D-Waves' quantum-classical hybrid framework, Fujitsu's quantum-inspired digital annealer, and Gurobi's state-of-the-art classical solver in solving a transport robot scheduling problem. This problem originates from an industrially relevant real-world scenario. We provide three different models for our problem following different design philosophies. In our benchmark, we focus on the solution quality and end-to-end runtime of different model and solver combinations. We find promising results for the digital annealer and some opportunities for the hybrid quantum annealer in direct comparison with Gurobi. Our study provides insights into the workflow for solving an application-oriented optimization problem with different strategies, and can be useful for evaluating the strengths and weaknesses of different approaches.
\end{abstract}


\section{Introduction}
\Ac{QC} is a field that has witnessed a rapid increase in interest and development over the past few decades since it was theoretically shown that quantum computers can provide an exponential speedup for certain tasks \autocite{deutsch1992rapid,grover1996fast,shor1994algorithms}. Translating this potential into a practically relevant quantum advantage, however, has proven to be a very challenging endeavor. Nevertheless, the emerging field is considered to have a highly disruptive potential for many domains, for example in machine learning \autocite{schuld2015introduction}, chemical simulations \autocite{cao2019quantum} and optimization \autocite{li2020}, the domain of this work. Due to the fact that optimization problems are of utmost importance also for industrial applications, we investigated a potential advantage of quantum and quantum-inspired technology for the so-called \ac{TRSP}, a real-world use-case in optimization that is derived from an industrial application of an automatized robot in a high-throughput laboratory. The optimization task is to plan a time-efficient schedule for the robot's movements as it transports chemical samples between a rack and multiple machines to conduct experiments. This is an NP-hard problem which for certain instances can be challenging to solve using classical computing techniques, and hence is an attractive candidate to search for an advantage with non-classical techniques. 

In our study, we compared the solution quality and runtime of different solvers on a large set of instances of the problem. As solvers, we considered \ac{LBQM} that makes use of the D-Wave quantum annealer \autocite{dwave2020}, \ac{FDA}\autocite{fujitsu2021}, \ac{FDAh}, as well as the industry-grade Gurobi solver \autocite{gurobi}. As a key element of this work, we provide three different models for the \ac{TRSP} that follow different design philosophies. This is justified by the different ways in which the problem task can be modelled and the inherent differences in the problem formulations that the solvers addressed can accept. \Ac{LBQM}, \ac{FDA} and \ac{FDAh} are restricted to a formulation as a \ac{QUBO}, whereas a \ac{MIP} with integer and float variables can be used by Gurobi, which makes a comparison of multiple formulations meaningful.

The \ac{TRSP} considered in this paper is a special combination of different scheduling problems that, to our knowledge, has not been considered before. Scheduling problems have been studied intensively for several decades and classical algorithms exist for numerous variants \autocite{scheduling,Pinedo2016}. Since most of the industry-relevant scheduling problems are NP-hard, these classical algorithms mainly consist of meta-heuristics or use general-purpose \ac{MIP} solvers, which basically solve the problem using a branch and bound approach with several additional improvements like cutting planes. In addition to classical algorithmic developments, a considerable amount of research has also been done in hardware-based parallel computing, especially in \ac{GPGPU} parallelization \autocite{chakroun, gpu_cdd}. The problem discussed in this work is an extension of the typical \ac{JSSP}, where the inclusion of a robot adds additional restrictions. More specifically, the studied scheduling problem falls into the category of robotic cell scheduling and \ac{AGV} scheduling problems. Most work on robotic cell scheduling deals with infinite cyclic schedules \autocite{dawande2007throughput}. This comprises polynomial-time algorithms and hardness results \autocite{Steiner2005}, \ac{MIP} techniques \autocite{phillips1976,brucker_transport,mip_robot_cell} and heuristic approaches \autocite{LIU2017188}. Many efficiently solvable and hard special cases have been identified \autocite{Shabtay2016} and heuristics have been proposed for some of the hard cases \autocite{Stern1990}. Those problems differ from our use case in one way or another. The problems considered by the above-cited papers allow, unlike our use case, that the jobs can wait at a machine after their completion before being picked up by the robot. Robotic cell scheduling problems without this possibility have been studied by \autocite{AGNETIS2000303,AgnetisPacciarelli2000}, whose problems differ from our, among others, in the considered objective function. Our objective function, the total job completion time, has been extensively studied for flow shop scheduling problems without a robot \autocite{Pinedo2016,Hall1996,Allahverdi2016,Roeck1984}, the latter of which shows that the no-wait variant is strongly NP-hard on two machines. Apart from the no-wait constraint, the problem considered in our work is characterized by the fact that jobs have to go to the last machine several times. Such settings are known as a re-entrant flow shops, for which \autocite{Jing2011} developed a heuristic algorithm. 

We are mainly interested in the performance of non-standard solution approaches using quantum or quantum-inspired solvers in this study. Because these solvers rely on heuristics, benchmarks for real-world applications are a highly relevant research topic. 
Most quantum optimization approaches fall into two major groups, one for gate-based hardware and one for annealing-based hardware \autocite{alexeev2021}. The majority of gate-based approaches to optimization use parameterized gates to find the ground state of a Hamiltonian related to the cost function of the optimization problem in a quantum-classical hybrid fashion, for example via the \ac{QAOA} \autocite{farhi2014,blekos2023}. 

Approaches based on quantum annealing also seek to find the ground state of a Hamiltonian, but by aiming for an adiabatic change from an initial state that can be easily prepared. In contrast to actual quantum computing devices, other classical software and hardware components are merely \emph{inspired} by quantum computing, for example \ac{FDA} \autocite{Aramon_2019} and \ac{TSB} \autocite{8892209}. Typically, optimization tasks for quantum solvers and the aforementioned quantum-inspired technologies are modeled as \ac{QUBO} problems \autocite{kochenberger2014}. An in-depth analysis of pure \ac{QUBO} comparison on four quantum and quantum-inspired solvers can be found in \autocite{Oshiyama2022}. In their work, the authors compare the solutions of a library of quadratic benchmark problems on the D-Wave quantum annealer, \ac{FDA}, and \ac{TSB} against each other.

\Ac{QC} has already been successfully used for optimization in various fields. For example, in \autocite{mizuno2023}, chemical reaction networks are optimized with quantum computing. In \autocite{Streif_2021}, it is shown that using the \ac{QAOA}, it is possible to beat some classical heuristic algorithms on the binary paint shop problem. %For optimization problems consisting of several constraints, 
However, some work has shown that the current circuit model algorithms are not always adequate enough to reach significant convergence required for a good solution \autocite{qc_knapsack}. Quantum annealing has proven to offer some advantage against the classical simulated annealing algorithm for a spin-glass problem, using D-Wave hardware \autocite{Raymond_2023}, but this is no conclusive evidence. In one of the more recent works on quantum annealing \autocite{schuetz}, the authors suggest a nature inspired hybrid quantum algorithm for robot trajectory optimization for PVC sealing in a real industrial setting. In \autocite{Ebadi_2022}, the authors present a solution to the \ac{MIS} problem using a Rydberg atom device, along with a claim of a possible super-linear quantum speed-up against classical simulated annealing. Other classical algorithms might still be superior to a quantum approach on current devices \autocite{Albash2018}.
Several works consider scheduling problems \autocite{yarkoni2021, carugno2022evaluating, tomasiewicz2020foundations}. In \autocite{agv_geitz}, an \ac{AGV} transportation problem using different classical and quantum approaches is studied and \autocite{ikeda2019} investigates a nurse scheduling problem with the usage of a quantum annealer.

The remaining manuscript is structured as follows. We provide a detailed description of the \ac{TRSP} and its mathematical modeling in \cref{sec:model}. In \cref{sec:setup}, we describe the design of our numerical study and list the problem instances and solvers that we use. The results of this study are presented in \cref{sec:results}. Finally, we conclude our study in \cref{sec:conclusions}. Detailed model descriptions, solver information, further information on the benchmark setup and instance lists are contained in the supplementary material (referenced by a preceding ``S'' to the label it is referring to).

\section{Transport Robot Scheduling Problem}
\label{sec:model}

In this section, we present a detailed explanation of the \ac{TRSP}, which is a real-world use case derived from one of BASF's high-throughput laboratories. This optimization problem is about finding the most time-efficient route of a transport robot tasked with moving chemical samples from one processing machine to another. 
In the following, we first provide a general description of the problem setup and then present different modeling approaches. These models build the foundation of the subsequent benchmarks. 

\subsection{Problem Description}

The laboratory we are modeling  consists of a \emph{sample rack} and three different processing machines: a \emph{water mixer}, a \emph{sample shaker} and a \emph{photo booth}. And, finally, the \emph{robot} itself that is tasked with carrying chemical samples from one place to another with the goal to conduct chemical experiments. Only the experimental plan (\ie, how each sample has to be processed in the laboratory) is predefined in advance, but not the specific order of the experiments. Initially, a certain number of samples is stored on the rack. Each of these samples needs to be first taken to the water mixer, then to the sample shaker. Once the sample shaking is completed, one or more photos have to be taken of each sample at the photo booth. Consecutive photos need to be taken after specific (\ie, predefined) time intervals, where the first photo of each sample has to be taken immediately after the shaking process. Finally, each sample has to be brought back to the rack. The processing times for different samples on the same machine can be different as specified by the experimental plan. We assume that each machine can only hold (and process) one single sample at any given time, or remain idle, and the processing steps cannot be interrupted before their completion. It is required that a machine starts processing a sample as soon as the sample is brought by the robot. Moreover, we assume that a sample has to be moved by the robot in-between two processing steps. Hence, a sample has to be lifted from a machine (and the machine is made available) as soon as it finishes processing. 

By definition, the robot requires exactly one time unit to move from any place to any other, with or without a sample, and picking up or dropping a sample does not require extra time. Like the machines, the robot can transport only a single sample at any given time or drive empty or remain idle. In particular, it is not possible that the robot places a sample at a machine and picks up another at the same time. 

The objective of this scheduling task is to minimize the \emph{sum of sample completion times}, \ie, the sum of the times when the samples arrive at the the rack after their last photo has been taken. The solution of this optimization problem is a sequence of tasks for the robot that yields an efficient laboratory operation. 

\subsection{Mathematical Modeling} \label{sec:modelling}
In our benchmark, we test three modeling approaches against each other. On the quantum and quantum-inspired side we consider a \ac{QUBO} formulation, whereas on the classical side we use two \ac{MIP} formulations. First, a so-called \emph{sequence model} and second, a so-called \emph{time-indexed model}. In the following, we first introduce the common terminology for all modeling approaches. Next, we shortly sketch the main features of each model. For a more detailed description, we refer to \ref*{sec:models}{}. The motivation for the development of multiple models is to carry out a comparison between the solutions obtained by the most suitable problem encoding for quantum and classical solvers. This ensures that we are comparing the best of both worlds (classical and quantum), and do not restrict ourselves to a model which is more suitable for quantum over classical computing.

\subsubsection{Common Terminology}
\begin{figure}[ht]
\centering
\includegraphics{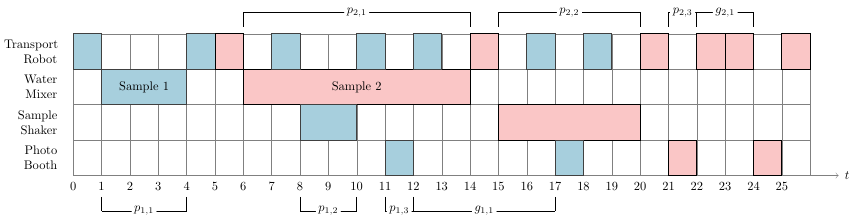}
\caption{An example Gantt chart of a robot transport scheduling problem with $N=2$ samples and $K=2$ photos.Tasks associated with sample one (two) are colored blue (red). When a sample is processed on one of the machines or carried by the robot in the time-frame $[t,t']$, a bar is drawn from $t$ to $t'$ in the respective row in a corresponding color. Empty movements of the robot are not drawn explicitly. For example, at time $t=13$ the robot is at the rack as sample~$1$ has been brought to the sample rack from $t=12$ to $13$. It takes one unit of time for the robot to travel from the rack to the water mixer to pick up sample~$2$ at $t=14$. From $t=22$ to $t=23$, the sample is brought from the photo booth to the rack and back, which is a consequence of the assumption that a sample has to be moved by the robot in-between two processing steps. The objective value of the depicted schedule is $19+26 = 45$.}
\label{fig:gantt}
\end{figure}

The processing machines are addressed by $M_1$ for the water mixer, $M_2$ for the sample shaker and $M_3$ for the photo booth. The scheduling time is discretized into time slots which all have length of one time unit. The transport robot takes one time unit for each operation that is either transportation or empty traversal between the machines and the rack. In this way, each transport robot scheduling problem is uniquely determined by the number of samples to be scheduled $N \geq 1$, the number of photos $K \geq 1$, which agrees for each sample $j \in \{1,\ldots,N\}$, the processing times $p_{j,1},p_{j,2}, p_{j,3} \in \mathbb{N}_{>0}$ for machines $M_1, M_2$ and $M_3$, which can vary for each sample $j \in \{1,\ldots,N\}$ and the time gaps $g_{j, k} \in \mathbb{N}_{\geq 2}$ to be kept between consecutive photos $k$ and $k+1$ for $k \in \{1,\ldots,K-1\}$, which also can vary for each sample $j \in \{1,\ldots,N\}$. As an example, \cref{fig:gantt} provides a feasible schedule in form of a Gantt chart to visualize these parameters.

\subsubsection{QUBO Model} \label{sec:qubo model}
A general \ac{QUBO} reads
\begin{equation} \label{eqn:qubo}
    \begin{aligned}
	\min_{x}\quad& x^{\top}\cdot Q \cdot x \\
	\text{s.t.}\quad& x \in \{0,1\}^n
\end{aligned}
\end{equation}
for some matrix $Q \in \mathbb R^{n \times n}$, where $x$ represents a vector of $n$ binary optimization variables. Two challenging properties of \acp{QUBO} must be taken into account in the modeling. Since only binary variables are allowed, this implies that other types of variables must be avoided, \ie a reformulation into a binary form is necessary. Second, the problem is unconstrained. This restriction can be overcome by using \emph{penalty terms}, which are quadratic functions in the model variables that evaluate to a positive value when the current assignment of values to the variables leads to an infeasible solution. Typically, the penalty terms are designed to yield $0$ if the corresponding solution is feasible, so that they do not contribute to the objective values of feasible solutions. More general information about \acp{QUBO} and their properties can be found, \eg, in \autocite{kochenberger2014,lucas2014,glover2018}. 

Our proposed \ac{QUBO} model for the \ac{TRSP} is based on the well-known starting time formulation (see \eg \autocite{carugno2022evaluating}) and can be written as
\begin{equation} \label{eqn:qubo-model}
    \begin{aligned}
	\min_{x}\quad& \rho_0 F(x) +  \sum_{i=1}^7 \rho_i P_i(x)\\
	\text{s.t.}\quad& x\in \{0,1\}^{n},
\end{aligned}
\end{equation}
where $F$ is the objective function and $P_1, \ldots, P_7$ denote the penalty functions and $\rho_0,\ldots,\rho_7 \in \mathbb{R}_{>0}$ are tunable parameters that have to be chosen such that the objective and penalty terms are suitably balanced. As in \cref{eqn:qubo}, $n$ represents the total number of binary optimization variables. These have a distinct meaning that can be identified with three indices. Specifically,
\begin{equation} \label{eqn:qubo-x}
	x_{j,m,t} :=
	\begin{cases}
		1, &\text{if sample $j$ starts processing on machine $M_m$ at time $t$}, \\
		0, &\text{otherwise}\;
	\end{cases}
\end{equation}
for all $j \in \{1,\ldots,N\}$, $m \in \{1,2,3\}$ and $t \in \{1,\ldots,T-1\}$. Here, $T$ denotes the time horizon, which is chosen in such a way that there is enough time to schedule all samples sequentially, implying that there is at least one feasible solution. It can be explicitly computed for each instance as described in \ref*{subsec:QuboModel}{}. In terms of \cref{fig:gantt}, one has, for example, $x_{1,1,1} = 1$ and $x_{1,2,8} = 1$.

The penalty terms for the \ac{QUBO} model have to be formulated using the binary optimization variables. This section only provides an example for such a term, a complete description can be found in \ref*{subsec:QuboModel}{}. Specifically, we consider here the constraint that each sample must access the machines $M_1$ and $M_2$ exactly once, which can be achieved by
\begin{equation}
P_1 := \sum_{j=1}^N  \sum_{m=1}^2 \left[ \left(\sum_{t=1}^{T-1} x_{j,m,t}\right) -1\right]^2 \;.
\end{equation}
This term evaluates to zero if and only if for each pair of sample $j$ and machine $M_m$, the variable $x_{j,m,t}$ is $1$ for precisely one time slot $t$. Since $P_1$ is bounded below by $0$ due to its quadratic nature, each local minimum of $P_1$ is a feasible solution \wrt the rule of machine access to $M_1$ and $M_2$. The other penalty terms can be formulated similarly.

Finally, the objective function $F$ sums up for each sample the time when the sample arrives at the rack after the entire scheduling process (``sum of sample completion times''). For example, the objective function in the case of \cref{fig:gantt} evaluates to $45$ time units. 

\subsubsection{MIP Models}

\Acp{MIP} have been used since the late 1950s as a tool for solving scheduling problems. It is not possible to model the disjunctive constraints resulting from the discrete ordering decisions only by means of starting time variables. Different types of binary variables have been proposed to achieve this. The main types are position variables~$x_{ijk}$ indicating if job $j$ is the $k$th job on machine~$i$ \autocite{Wagner1959}, linear ordering variables~$\delta_{ijk}$ deciding if job~$j$ is processed before job~$k$ on machine~$i$ \autocite{Manne1960} and time-indexed variables~$x_{ijt}$ specifying that job~$j$ is started (or processed or completed) on machine~$i$ at time~$t$ \autocite{Bowman1959,Pritsker1969}. \autocite{Ku2016} compared these three approaches experimentally for a job shop scheduling problem.

Due to the powerful nature of (mixed) integer programming in contrast to the restrictive nature of the \ac{QUBO} models, we provide two \ac{MIP} models to be solved using Gurobi, where we follow two state-of-the-art approaches  for formulating scheduling problems as MIPs\autocite{Pinedo2016}. The first one, in the following named \emph{sequence model}, makes use of continuous start time and binary linear ordering variables. The second model, called the \emph{time-indexed model}, is restricted to a binary formulation comparable to the \ac{QUBO} model, where we make use of time-indexed variables. The latter provides a model with a natural vicinity to the \ac{QUBO} formulation whereas the sequence model exploits the features of MIP formulations. In this sense we provide a baseline from two different angles, one for each solution approach. 

\subsubsection{MIP: Sequence Model}

In the sequence model, we model sequences of \emph{events} that affect the behavior of the transport robot with respect to the machines and the photos of a sample. We define the \emph{set of events} as
\begin{equation}
E := \bigl\{(j, i, a) \mid j \in \{1,\ldots, N\},\, i \in \{1, \ldots, 2 + K\},\, a \in \{0,1\}\bigr\}.
\end{equation}
An event $e = (j,i,0)$ represents either that a sample $j$ is placed on machine $M_i$ for $i \in \{1,2\}$ or to the $(i-2)$th photo shoot for $i > 2$, an event~$(j,i,1)$ corresponds to picking it up again. For each event~$e \in E$ we define an optimization variable $\tau_e \in \mathbb{R}_{\geq 0}$ to model the time for event $e$ to happen. In terms of \cref{fig:gantt}, we have, for example, $\tau_{(1,1,0)} = 1$ and $\tau_{(1,1,1)}=4$. A simple formulation can be achieved by additionally introducing a binary variable for each pair $e,f \in E$, $e \neq f$ of events that indicates if $e$ occurs before $f$. We reduce the size of the model by exploiting the fact that the ordering of some events is fixed or coupled. For example, we do not need a variable that specifies the order in which a given sample is brought to the water mixer and to the sample shaker. This leads to three sets of linear ordering variables that can be found in \ref*{subsec:SequenceModel}{}, as well as the various  constraints to ensure feasibility. The objective function (\ie, the sum of the sample completion times) can be easily expressed using the variables~$\tau_e$ corresponding to events when a sample is picked up from the last photo.

\subsubsection{MIP: Time-Indexed Model}

The second constrained model makes use of discrete time-indexed variables similar to the \ac{QUBO} model from \ref*{subsec:QuboModel}{}. In this formulation, we model the behavior of the transport robot by defining certain routes a sample can be transported along, which include those from the rack to all machines and back or movements between subsequent machines. The numbering of the moves is shown in \ref*{fig:routestimeindexed}{}.

As the model name implies, we have, given a discrete time horizon $T \in \mathbb{N}_{>0}$, binary variables to model when each sample takes which route as

\begin{equation}
	y_{j, r,t} :=
	\begin{cases}
		1, \text{ if sample $j$ is transported by the robot on route $r$ during the time $(t, t+1)$} \;, \\
		0, \text{ otherwise}\;
	\end{cases}
\end{equation} for all $j \in \{1,\ldots,N\}$, $r \in \{1,\ldots, 8\}$ and $t \in \{0,\ldots,T-1\}$. In terms of the Gantt chart from \cref{fig:gantt}, this would imply $y_{1,1,0} = 1$, $y_{1,2,4} = 1$, $y_{2,1,5} =1$ and so on. The time horizon $T$ is defined as for the QUBO model, see \ref*{eq:QuboTimeHorizon}{}.

The constraints of the model are similar to the penalty terms of the \ac{QUBO} Model and are listed in \ref*{subsec:TimeIndexedModel}{}. The objective function (\ie, the sum of the sample completion times) is defined in terms of the ancilla optimization variables $z_j$ for $j\in\{1,\dots,N\}$, that are bounded below by the arrival time of sample $j$ at the rack after the schedule has finished.

\section{Benchmark Setup}
\label{sec:setup}

In the present section, we describe the design of the benchmark. We start with an outline of the considered problem instances that are listed in more detail in \ref*{sec:benchmark library}{}. Subsequently, we describe the three different commercial technologies that we use.

\subsection{Instances} \label{sec:instances}

To set the stage for our benchmark, we specify \num{260} test instances of our optimization problem of interest, each defined by a different set of parameters. Specifically, each instance is uniquely determined by the number of samples $N$, the number of photos $K$, the gaps $g_{j,k}$ between subsequent photos $k$ and $k+1$ for $k \in \{1, \ldots, K-1\}$ and $j \in \{1, \ldots, N\}$, and, finally, the processing times $p_{j,1}, p_{j,2}, p_{j,3}$ of the water mixer, sample shaker and photo booth, respectively, as explained in \cref{sec:modelling}. For the sake of simplicity, the processing time of the photo booth agrees for all samples of the same instance, that is $p_{j,3} := p_3$ for all $j \in \{1,\ldots,N\}$.

In \ref*{sec:benchmark library}{}, we describe the algorithm that was used to generate parameter sets for the benchmark instances. Since the resulting instances span a wide range of complexity, we divide the resulting benchmark library into two parts, where each part is defined by the number of binary variables in the corresponding \ac{QUBO} formulation from \cref{sec:qubo model} as explained in \ref*{subsec:QuboModel}{} in more detail. The first part, which we call \emph{library of minor instances}, contains all \num{161} instances that have at least \num{2071} and at most \num{8080} binary variables. The second part, which we call \emph{library of major instances}, contains the remaining \num{99} instances with at least \num{10822} and at most \num{22692} binary variables. The reason for that specific division is that \num{8192} is the maximal amount of variables that can be solved directly on Fujitsu's digital annealer. 

We collect groups of instances $(N,K)$ that have the same number of samples and photos as shown in \ref*{fig:NumVariablesPerBox}{}, \ie, within those groups the leftover parameters $p_{j,m}$ and $g_{j, k}$ for $j \in \{1,\ldots,N\}, m \in \{1,2,3\}$ and $k \in \{1,\ldots,K-1\}$ may vary. These groups can be understood as a collection of ``similar'' \acp{TRSP} in the sense that the complexity of the tasks to be solved is comparable. However, some instances may still be easier or more difficult to solve than others in practice. This grouping approach allows us to consider statistical metrics over several instances when we compare models and solvers. Moreover, it allows us to estimate the scaling behavior of different solution approaches. In \ref*{sec:benchmark library}{}, we list how many instances each group contains.

\subsection{Quantum and classical solvers} \label{sec:solvers}

In our benchmark, we solve the generated instances with a selection of model and solver combinations with the main goal to assess the performance of quantum and quantum-inspired technology. Specifically, we consider three solver candidates:
\begin{enumerate}
\item Gurobi: As a baseline, we use the branch and bound algorithm of Gurobi, which is a state-of-the-art mathematical programming solver running on classical hardware \autocite{gurobi}. In summary, it relies on an implicit enumeration that allows the original problem to be split into smaller sub-problems using a decision tree. The use of lower bounds derived from \ac{LP} relaxations allows for a reduction of the search space. Gurobi is an all-purpose solver that can in principle solve the proposed optimization problems to a guaranteed optimality in a deterministic fashion (given sufficient time). In this work we utilized the cloud based service of Gurobi solver, which ran on a Intel(R) Xeon(R) Platinum 8275CL CPU (3.00 GHz with 8 physical cores).
\item \acf{LBQM}: D-Wave provides cloud-based access to their adiabatic quantum computers with over \num{5000} qubits \autocite{dwave2020}. By design, their hardware is specifically tailored to solve \acp{QUBO}. To this end, the \ac{QUBO} is encoded in an Hamiltonian such that each optimization variable is represented by one qubit \autocite{qubo_embedding} and the ground state corresponds to the optimal solution. The quantum annealing mechanism aims to find the ground state by performing a suitable time evolution of the quantum system with a subsequent measurement of all qubits to reveal the optimal solution.\par 
The D-Wave hardware has only limited connectivity, which means that each qubit can only interact with a certain number of other qubits. This limitation restricts the correlations between optimization variables that can be represented by the Hamiltonian. Finding a suitable representation with these constraints is an NP-hard problem \autocite{lobe2022} that has to be solved classically to configure the quantum annealer for a certain problem. In practice, the quantum annealer can typically only be used for \acp{QUBO} with much less than \num{5000} optimization variables.\par
For this reason, D-Wave also provides a hybrid software framework \ac{LBQM}, which is a black-box algorithm for \acp{BQM} that runs on both classical and quantum annealing hardware. It allows larger optimization problems that are too big for the quantum hardware to be handled by presenting only parts of the original problem to the quantum annealer. However, the exact mode of operation of \ac{LBQM} is not publicly available. In this study, we use only the quantum annealer in a hybrid fashion via \ac{LBQM}. The quantum machine used in the hybrid framework is the \emph{D-Wave Advantage System 4.1} and the region \emph{na-west-1}. We choose to use a constant number of \num{1000} samples (or readouts) for all evaluations and use default settings for all parameters.
\item \acf{FDA} and \acf{FDAh}: The digital annealer from Fujitsu can be considered as a quantum-inspired algorithm that runs on dedicated (classical) hardware \autocite{Aramon_2019} and can be accessed using a cloud service. It is based on simulated annealing \autocite{Kirkpatrick1983, Cerny1985} with two major differences. Firstly, the utilization of an efficient parallel-trial scheme to exploit the parallelization capabilities of the hardware and, secondly, a dynamic escape mechanism to avoid locally optimal solutions. The detailed hardware specifications are confidential. The solver supports \acp{QUBO} with up to \num{8192} variables.\par
In addition, the hybrid solver \ac{FDAh} is provided to solver bigger problem instances by utilizing both dedicated and classical hardware \autocite{fujitsu2021} similar to D-Wave's \ac{LBQM}. In this study, we use both \ac{FDA} and \ac{FDAh}. Both solvers require a set of parameters that specify how the annealing is done, which also include the number of repetitions and parallel runs on the chip. The specific parameters we used for \ac{FDA} and \ac{FDAh} are provided in \ref*{sec:solver settings}{}.
\end{enumerate}

In a small pre-study, we excluded a few other solvers; see \ref*{sec:prestudy}{}. The main scope of the paper is to benchmark the performance of quantum-hybrid and quantum-inspired technologies on the \ac{TRSP} on a high level against an all-purpose solver with an out-of-the-box performance. In this sense, we also exclude meta-heuristics that are tailor-made to the problem as well.

Each instance can be modelled with each of the three modeling approaches from \cref{sec:model}. However, not all solvers are applicable to all problem formulations and all instances. The \ac{MIP} sequence model is solved with Gurobi for all instances. The time-indexed model is solved with Gurobi only for the minor instances. The \ac{QUBO} model is solved with \ac{LBQM} and \ac{FDA} for minor instances. For major instances, the \ac{QUBO} model is only solved with \ac{FDAh}.

\begin{figure}[!ht]
\begin{center}
\includegraphics[]{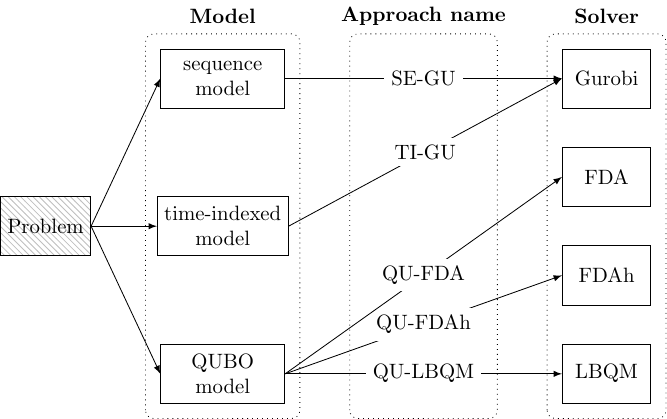}
\caption{Summary of model (see \cref{sec:modelling}) and solver (see \cref{sec:solvers}) combinations for the benchmarks.}\label{fig:approaches}
\end{center}
\end{figure}

We call each valid model and solver combination an \emph{approach} and use a unique name to refer to it. Summarized, we consider \ac{SG}, \ac{TG}, \ac{QLBQM}, \ac{QFDA} and \ac{QFDAh}. An overview over all approaches is shown in \cref{fig:approaches}.

For all problems, we prescribe a runtime limit of \num{3600} seconds for Gurobi. This limit was determined on a heuristic basis, since initial experiments have shown that Gurobi can solve the considered problem instances on this time scale with a practically relevant quality. This time limit exceeds the runtimes of \ac{LBQM}, \ac{FDA} and \ac{FDAh} by far to provide Gurobi enough time to return solutions that are suitable for a relative comparison (see \cref{fig:RelativeRuntimeComputation}).

Both \ac{LBQM} and \ac{FDA} also require a time limit for each run, which scales with the problem size in the \ac{QUBO} formulation as follows. The time limit for \ac{LBQM} is set to be $\min\{100, 1.5 \cdot \frac{n}{100}\}$ seconds, where $n$ is the number of variables in the \ac{QUBO} formulation for the minor instances. The runtime of the digital annealer is implicitly set with the \emph{steps} parameter, where each step taken in the annealing process takes a constant amount of time. We set the number of steps to be $\num{1e7}$ for the instances with $\num{2071} \leq n \leq \num{4096}$, $\num{5e7}$ for the ones with $\num{4096} < n \leq \num{6000}$ and $\num{1e8}$ for the instances with $\num{6000} < n \leq \num{8080}$ variables in the \ac{QUBO} formulation. Lastly the major instances computed with the hybrid framework \ac{FDAh} based on the digital annealer require a time limit as well. For this we distributed the available time of \num{5} hours to the instances, correspondingly to their number of variables. This computes approximatively as $n \cdot \num{0.0117}$ seconds where $n$ is the number of variables in the \ac{QUBO} formulation.

The benchmark setup is summarized in \cref{tab:BenchmarkSetups}, where we recall the approaches from \cref{fig:approaches}. The table also contains the values of the \ac{QUBO} parameters $\rho_0,\ldots, \rho_7$ from \ref*{eq:QuboModel}{} that were chosen for \ac{LBQM}, \ac{FDA} and \ac{FDAh}, respectively. The choice was made according to previous experiments with smaller problem instances. For this purpose, a typical strategy is to iteratively increase the parameter $\rho_i$ if the corresponding penalty term $P_i$ is non-vanishing. Additionally, one needs to make sure that the parameter $\rho_0$ for the target function is set such that it is not in favor to violate penalty terms and a good optimization is achieved.

Some solutions of the library of minor instances have not been solved to feasibility by \ac{LBQM}, \ie, the solution vector returned does not translate to a feasible schedule of the \ac{TRSP}. Those instances can be identified by having an objective value of at least $10^4$, which is the minimum of the penalty parameters chosen for the \ac{QUBO} model according to \cref{tab:BenchmarkSetups}. This can be seen as follows: the parameters of the library of minor instances are bounded as $N \leq 9$, $K \leq 4$, $p_{3,j} \leq 3, p_{1,j} \leq 8, p_{2,j} \leq 4, g_{1,j} \leq 5, g_{2,j} \leq 12$ and $g_{3,j} \leq 24$ for $j=1,\ldots,N$. Using those upper bounds we compute a maximal time horizon of $T = 648$ time units for those instances. It follows that the sum of sample completion times is bounded above by $9 \cdot 648 = 5832 < 10^4$, \ie, a solution to an instance of the library of minor instances is feasible if and only if it has an objective value below $10^4$. Of course this does neither apply to the library of major instances nor to the solutions of \ac{FDA} or \ac{FDAh} as they have lower penalty parameters due to prestudies with the smallest instances. In a general setup a way to identify infeasible solutions is to store the penalty term $\sum_{i=1}^7 P_i(x)$ and evaluate the solution with it. The solution is feasible in this case if and only if the penalty term evaluates to $0$ on it.

\begin{table}[!ht]
    \begin{center}
        \caption{Benchmark setup: Summary of problem instances from \cref{sec:instances} and solvers from \cref{sec:solvers} for the optimization problems (or models) from \cref{sec:model}.}\label{tab:BenchmarkSetups}
        \begin{tabular}{lcc}
            \toprule
            Property & Minor instances & Major instances \\ \midrule
            Number of instances & $\num{161}$ & $\num{99}$ \\
            Number of variables $(n)$ & $\num{2071} \text{ to } \num{8080}$ & $\num{10822} \text{ to } \num{22692}$ \\ \midrule
            Approach (\cf \cref{fig:approaches}) &  Used for minor instances & Used for major instances \\ \midrule
            \ac{SG} & \checkmark &  \checkmark \\ 
            \ac{TG} & \checkmark & \xmark\\ 
            \ac{QLBQM} & \checkmark & \xmark \\ 
            \ac{QFDA} & \checkmark & \xmark \\
            \ac{QFDAh} & \xmark & \checkmark 
            \\\midrule
            Approach (\cf \cref{fig:approaches}) &  Minor instance limit & Major instance limit  \\ \midrule
            \ac{SG} & \SI{3600}{\second} &  \SI{3600}{\second} \\ 
            \ac{TG} & \SI{3600}{\second} & ---\\ 
            \ac{QLBQM} & $\min\{\num{100}, \num{1.5} \cdot \frac{n}{\num{100}}\} \cdot \SI{1}{\second}$ & --- \\ 
            \ac{QFDA} & 
            $ 
                \begin{cases} 
                    \num[print-unity-mantissa=true]{1e7} \text{ iterations,} &\num{2071} \leq n \leq \num{4096} \\
                    \num[print-unity-mantissa=true]{5e7} \text{ iterations,} &\num{4096} < n \leq \num{6000} \\
                    \num[print-unity-mantissa=true]{1e8} \text{ iterations,} &\num{6000} < n \leq \num{8080}
                \end{cases}
            $ & --- \\ 
            \ac{QFDAh} & --- & $n \cdot \SI{0.0117}{\second}$ \\
            \midrule 
            Solver & \multicolumn{2}{c}{\Ac{QUBO} parameters from \ref*{eq:QuboModel}{}}  \\
            \midrule
            \ac{LBQM} & \multicolumn{2}{c}{$\rho_0= \num{1}, \rho_1= \num{30000}, \rho_2= \rho_3= \rho_4= \rho_5= \rho_7= \num{10000}, \rho_6= \num{15000}$} \\
            \ac{FDA} & \multicolumn{2}{c}{$\rho_0= \num{1000}, \rho_1= \num{4000}, \rho_2= \rho_3= \num{1000}, \rho_4= \rho_5= \rho_6= \rho_7= \num{1500}$} \\
            \ac{FDAh} & \multicolumn{2}{c}{$\rho_0= \num{1000}, \rho_1= \num{2000}, \rho_2= \rho_3= \num{500}, \rho_4= \rho_5= \rho_6= \rho_7= \num{750}$} \\
             \bottomrule \\
        \end{tabular} 
    \end{center}
\end{table}

\section{Benchmark Results}
\label{sec:results}

In the current section, we present the results of our previously described benchmark, which is summarized in \cref{tab:BenchmarkSetups}. For this purpose, we first show the results for the minor instances and subsequently the results for the major instances. 

\subsection{Results for Minor Instances}

\begin{figure}[!htp]
    \centering
    \begin{subfigure}[b]{\textwidth}
    \centering
        \includegraphics{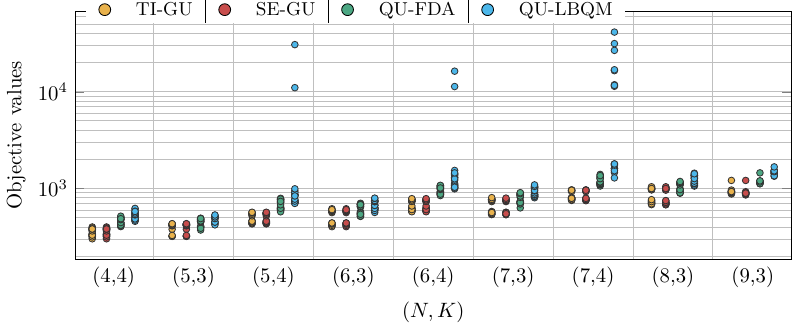}
        \caption{Objective values}
        \label{fig:BenchmarkAObjective}
    \end{subfigure}
    \\[.65cm]
    \begin{subfigure}[b]{\textwidth}
        \centering
        \includegraphics{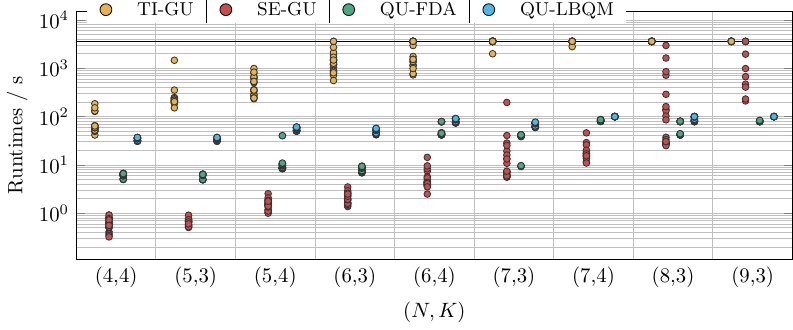}
        \caption{Runtimes}
        \label{fig:BenchmarkARuntime}
    \end{subfigure}
    \caption{Benchmark results for minor instances as scatter plots. The results are grouped into sets of instances $(N,K)$ with the same number of samples $N$ and photos $K$. A horizontal line marks the upper time limit of \SI{3600}{\second} for Gurobi in \cref{fig:BenchmarkARuntime}. Some instances have not been solved to feasibility by \ac{QLBQM}, as indicated by the peaks above $10^4$ in \cref{fig:BenchmarkAObjective}. Abbreviations according to \cref{fig:approaches}.}
    \label{fig:BenchmarkAObjectiveAndRuntime}
\end{figure}

In \cref{fig:BenchmarkAObjectiveAndRuntime}, we show the objective values and runtimes of several approaches as scatter plots. All runtimes are end-to-end runtimes, that is, we consider the entire evaluation pipeline, beginning with the submission of the problem to the solver and ending with the return of a solution, including potential network delays. The programmatic construction of the optimization problem for the \ac{API} of the solver based on the instance data is not part of the runtime.

From \cref{fig:BenchmarkAObjective}, we can observe that both the \ac{SG} and \ac{TG} solutions reach a better objective value than the solutions from \ac{QLBQM} and \ac{QFDA}. When comparing objective values, it has to be taken into account that the \ac{QUBO} model objective, \cref{eqn:qubo-model}, also includes penalty terms, which become positive for infeasible solutions and therefore increase the objective value accordingly. Specifically, we find that only \ac{QLBQM} yields infeasible solutions for some instances, whereas all other approaches yield feasible solutions (\ac{SG} and \ac{TG} solutions are by definition always feasible). For our analysis, we include both feasible and infeasible solutions. By performing a Welch t-test \autocite{welch1947}, we find that the means of the results from both \ac{SG} and \ac{TG} are lower than the means of the \ac{QFDA} and \ac{QLBQM} results with a statistical significance of over $99 \%$, respectively. The same holds for the \ac{QFDA} objective values in comparison to \ac{QLBQM}.

\begin{figure}[!htp]
    \begin{center}
        \includegraphics[]{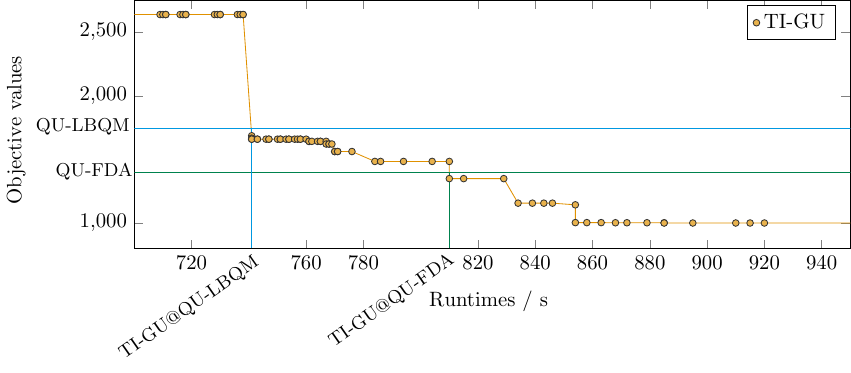}
        \caption{Visualization of the relative runtime of \ac{TG} \wrt \ac{QLBQM} and \ac{QFDA}, denoted by \ac{TG-QLBQM} and \ac{TG-QFDA}, respectively. Here, we consider the example instance $(7,4,3)(3)$; see supplementary material. The orange dots (connected by lines for better visualization) mark the resulting objective values of \ac{TG} at the corresponding time steps. The horizontal upper, blue and lower, green line mark the final objective value of \ac{QFDA} and \ac{QLBQM}, respectively, on the same instance. The blue and green lines intersect with the orange lines at some point. The time coordinate of the next lower \ac{TG} objective value after this intersection represents the relative runtime of \ac{TG} \wrt the solver, which is marked as a vertical line in the corresponding color. In other words, the relative runtime represents how long \ac{TG} has to run until it reaches an objective value that is at least as good as the result from \ac{QLBQM} or \ac{QFDA}, respectively.}
        \label{fig:RelativeRuntimeComputation}
    \end{center}
\end{figure}

On the other hand, according to \cref{fig:BenchmarkARuntime}, the computation time for \ac{TG} and for some instances of \ac{SG} exceed the computation time of \ac{QLBQM} and \ac{QFDA}. Since \ac{MIP} solvers typically spend a lot of time proving that a solution is optimal, we are also interested in the time taken by Gurobi (for both \ac{SG} and \ac{TG}) to find solutions of the same quality as those obtained from \ac{QLBQM} or \ac{QFDA}. Hence, we perform an additional analysis of the iterative solver progress of each Gurobi run and look for the earliest computation time at which Gurobi has reached an objective value that is less than or equal to the corresponding objective value returned by the competing solvers for the same instance. We call this earliest computation time the \emph{relative runtime}. Specifically, we consider \ac{TG-QLBQM}, \ac{SG-QLBQM}, \ac{TG-QFDA} and \ac{SG-QFDA}. In the special case that Gurobi is not able to find an objective value of the desired quality within its limit of \num{3600} seconds (which only occurs for some major instances), this time limit is used in place of the earliest computation time. Exemplarily, we consider a specific instance to visualize \ac{TG-QLBQM} and \ac{TG-QFDA} in \cref{fig:RelativeRuntimeComputation}. 

\begin{figure}[!htp]
    \begin{subfigure}[b]{\textwidth}
        \begin{center}
            \includegraphics[]{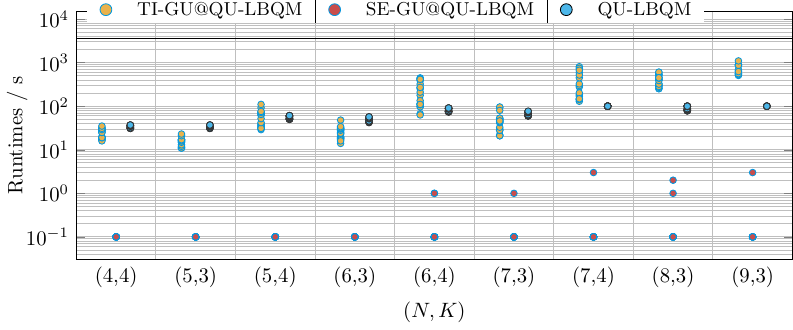}
            \caption{Runtime comparison for \ac{TG-QLBQM}, \ac{SG-QLBQM} and \ac{QLBQM} from \cref{fig:BenchmarkARuntime}.}
            \label{fig:BenchmarkALeapRelativeRuntime}
        \end{center}
    \end{subfigure}
    \\[.65cm]
    \begin{subfigure}[b]{\textwidth}
        \begin{center}
            \includegraphics[]{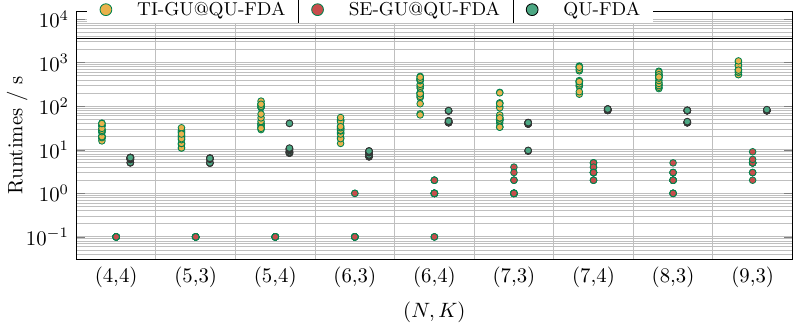}
            \caption{Runtime comparison for \ac{TG-QFDA}, \ac{SG-QFDA} and \ac{QFDA} from \cref{fig:BenchmarkARuntime}.}
            \label{fig:BenchmarkAFujitsuRelativeRuntime}
        \end{center}
    \end{subfigure}
    \caption{Benchmark results for minor instances as scatter plots. We show the relative runtimes of \ac{TG} and \ac{SG} \wrt \ac{QLBQM} and \ac{QFDA}, denoted by \ac{TG-QLBQM}, \ac{TG-QFDA}, \ac{SG-QLBQM} and \ac{SG-QFDA}, respectively. The results are grouped into sets of instances $(N,K)$ in analogy to \cref{fig:BenchmarkAObjectiveAndRuntime}. See \cref{fig:RelativeRuntimeComputation} for an example of the relative runtime computation. Abbreviations according to \cref{fig:approaches}.}
    \label{fig:BenchmarkARelativeRuntimesBoth}
\end{figure}

The results of this analysis are shown in \cref{fig:BenchmarkARelativeRuntimesBoth}. This plot shows that \ac{QLBQM} is not able to compete with \ac{SG}. All problems from the first $4$ out of $9$ instance groups have been solved with \ac{SG} in under $1$ second while the remaining instances in less than $10$ seconds, whereas the \ac{QLBQM} runtimes range between $50$ and $100$ seconds. However, \ac{LBQM} finds a comparable solution faster than \ac{TG} for most problems with $6$ or more samples and remains competitive for smaller problems. A Welch-t test confirms that the mean of \ac{TG} runtime is larger than the one of \ac{QLBQM} runtime with a significance over $99 \%$. 

Furthermore, \cref{fig:BenchmarkAFujitsuRelativeRuntime} shows that \ac{QFDA} is outperformed by \ac{SG} as well. Analogous to \cref{fig:BenchmarkALeapRelativeRuntime}, the instances in groups $(4,4), (5,3), (5,4)$ and $(6,3)$ have been solved by \ac{SG} in $1$ second or less. But in contrast to \cref{fig:BenchmarkALeapRelativeRuntime}, the other groups have their median between $1$~second and $10$~seconds, \ie, which reflects that the target objectives from \ac{QFDA} are lower than those from \ac{QLBQM} (see \cref{fig:BenchmarkAObjective}). Nonetheless, the time taken for \ac{SG} to reach the solution quality of \ac{QFDA} is $10$ to $100$ times smaller. Regarding \ac{TG}, \ac{QFDA} finds a comparable solution almost always faster with a few exceptions.

\subsection{Results for Major Instances}
\label{subsec:resultsmajor}

\begin{figure}[!htp]
\begin{subfigure}[b]{\textwidth}
\begin{center}
\includegraphics[]{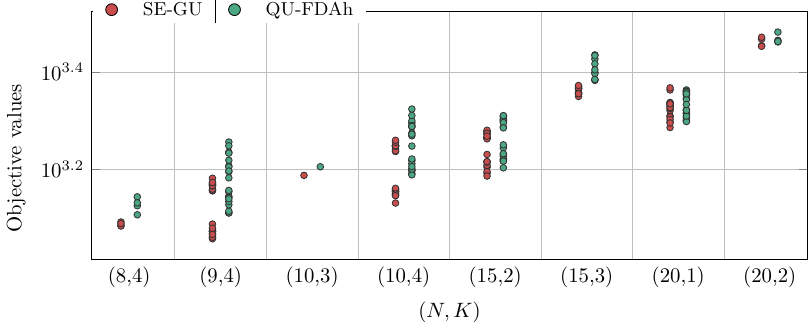}
\caption{Objective values}
\label{fig:BenchmarkBObjective}
\end{center}
\end{subfigure}
\\[.65cm]
\begin{subfigure}[b]{\textwidth}
\begin{center}
\includegraphics[]{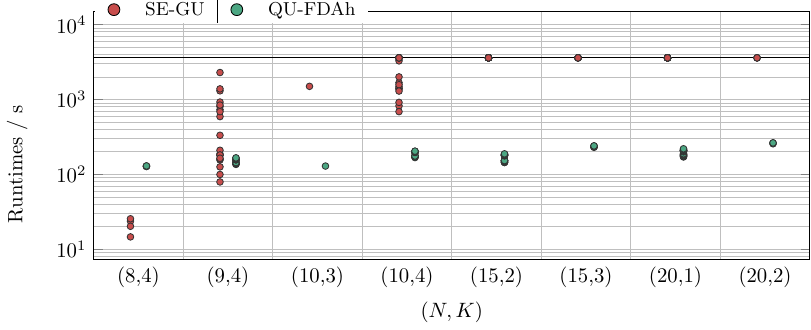}
\caption{Runtimes}
\label{fig:BenchmarkBRuntime}
\end{center}
\end{subfigure}
\caption{Benchmark results for major instances as scatter plots. The results are grouped into sets of instances $(N,K)$ as for previous the plots. Abbreviations according to \cref{fig:approaches}.
}
\label{fig:BenchmarkBObjectiveAndRuntime}
\end{figure}

The results for major instances are presented in analogy to the results for minor instances from the previous section. In \cref{fig:BenchmarkBObjectiveAndRuntime}, we show the runtime and the target value of the solvers on the corresponding models as scatter plots.

The objective values of \ac{QFDAh} are worse than the ones of \ac{SG} with a significance of over $97\%$, but \cref{fig:BenchmarkBRuntime} shows that the runtime of \ac{SG} increases strictly until it reaches the upper bound for the computation time of \num{3600} seconds, which happens for ca.~$15$ samples. On the other hand, the computation time of \ac{QFDAh} ranges between \num{120} and \num{300} seconds, where only a slight increase can be seen.

\begin{figure}[!htp]
\begin{center}
\includegraphics[]{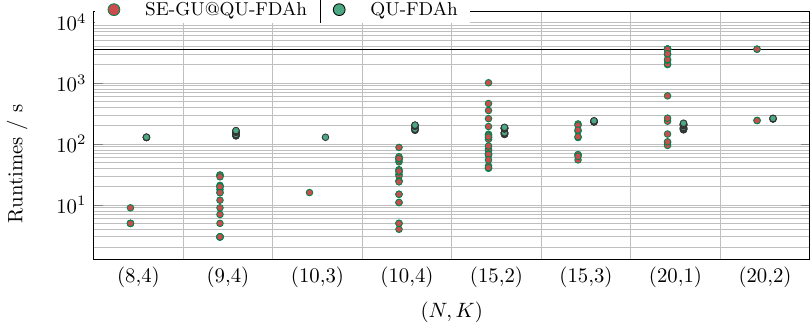}
\caption{Benchmark results for major instances as scatter plots. We show the relative runtime of \ac{SG} \wrt \ac{QFDAh}, denoted by \ac{SG-QFDAh}, in analogy to \cref{fig:BenchmarkARelativeRuntimesBoth}. We also show the runtime of \ac{QFDAh} from \cref{fig:BenchmarkBRuntime}. The results are grouped into sets of instances $(N,K)$ as for previous plots. Abbreviations according to \cref{fig:approaches}.
}\label{fig:BenchmarkBFujitsuRelativeRuntime}
\end{center}
\end{figure}

Analogously to \cref{fig:BenchmarkAFujitsuRelativeRuntime}, we evaluate the earliest computation times of \ac{SG} model to reach objective values equal to or lower than the objective values obtained from \ac{QFDAh}, denoted by \ac{SG-QFDAh}. The results are shown in \cref{fig:BenchmarkBFujitsuRelativeRuntime}.

In \cref{fig:BenchmarkBFujitsuRelativeRuntime}, a strictly increasing computation time can be seen for  \ac{SG}, whereas the \ac{QFDAh} runtime remains almost constant. For the biggest instances with $N=20$ samples, \ac{QFDAh} has a clear advantage with respect to the computation time, whereas it is competitive to \ac{SG} for the instances with $15$ samples. In this sense \ac{QFDAh} finds a solution of comparable quality much faster for problems with $20$ samples than \ac{SG} and the latter was not able to prove optimality for some of the instances with $20$ samples. A Welch t-test confirms with a significance of over $99 \%$ that the \ac{QFDAh} mean is lower than the \ac{SG-QFDAh} mean.

\section{Conclusion and Outlook}
\label{sec:conclusions}
\acresetall

This paper presents a thorough benchmarking of an industrially relevant use case of combinatorial optimization, the \ac{TRSP} with the goal to achieve a time-optimal robot schedule, as motivated by a BASF high-throughput laboratory. We solve a large set of instances for this optimization problem with varying difficulty using three commercially available solvers: (i) the D-Wave’s hybrid Leap framework, (ii) the quantum-inspired Fujitsu digital annealer and (iii) the classical state-of-the-art solver Gurobi. To this end, we develop several mathematical models: a \ac{QUBO} model for the quantum and digital annealer and two different \ac{MIP} models for Gurobi, which we call time-indexed and sequence model, respectively. Modeling the same problem in different, solver-specific forms helps us to optimally assess the capabilities of each solver. In total, we compare five different approaches (\ie, model and solver combinations as sketched in \cref{fig:approaches}): (i) \ac{TG}, (ii) \ac{SG}, (iii) \ac{QLBQM}, (iv) \ac{QFDA} and (v) \ac{QFDAh}. For our performance study, we separated all problem instances into two groups. First, the \emph{minor instances} with problems less than \num{10000} binary variables in the \ac{QUBO} formulation and, second, the \emph{major instances} with problems with more than \num{10000} and up to \num{22000} variables. For practical reasons, we only solve the minor instances with \ac{SG}, \ac{TG}, \ac{QLBQM} and \ac{QFDA}, whereas the major instances are only solved with \ac{SG} and \ac{QFDAh}, respectively.

Our benchmark reveals insights both regarding the objective values of the optimization problem (\ie, the sum of sample completion times) as well as the end-to-end runtimes for the considered approaches. Regarding the objective values, we observe for minor instances that \ac{SG} and \ac{TG} give similar results, outperforming \ac{QFDA}, which in turn outperforms \ac{QLBQM} (\cf \cref{fig:BenchmarkAObjective}). For major instances, \ac{SG} outperforms \ac{QFDAh} (\cf \cref{fig:BenchmarkBObjective}). Regarding the runtime, we find that for smaller instances \ac{TG} takes the highest time and \ac{SG} takes mostly the lowest. Between these two extremes, \ac{QFDA} and \ac{QLBQM} take about the same amount of time (\cf \cref{fig:BenchmarkARuntime}). However, the runtime of \ac{SG} significantly increases with increasing instance complexity. This same observation continues for the large instances, for which the runtime of \ac{SG} is mostly larger than that of \ac{QFDAh} (\cf \cref{fig:BenchmarkBRuntime}).

To get further insights into the relationship between objective value and runtime, we also studied the relative runtime of Gurobi, that is the time that Gurobi took to find an objective value that is at least as good as the final result from another approach. For minor instances, we find that the relative runtimes of \ac{SG} \wrt \ac{QLBQM} and \ac{QFDA}, respectively, are strictly lower than the runtimes of \ac{QLBQM} and \ac{QFDA}, \ie, Gurobi found solutions of comparable quality faster than the quantum and quantum-inspired approaches (\cf \cref{fig:BenchmarkALeapRelativeRuntime,fig:BenchmarkAFujitsuRelativeRuntime}). This is not surprising since \ac{SG} tended to find better objectives in shorter time. For major instances, the relative runtimes of \ac{SG} \wrt \ac{QFDAh} increase significantly with increasing instance complexity and clearly exceed the runtime of \ac{QFDAh} for the biggest instances (\cf \cref{fig:BenchmarkBFujitsuRelativeRuntime}). Thus, \ac{QFDAh} shows an advantage on some bigger instances. Although the resulting objective values of \ac{QFDAh} were not optimal, the approach shows a clear advantage on some bigger instances when compared to \ac{SG} on a similar time scale.

Our benchmark spans instances of different scales and therefore allows qualitative estimation of the scaling behavior of different approaches. Specifically, we observe that \ac{TG} and \ac{SG} show a runtime that scales exponentially with the instance complexity (as estimated by the number of samples and photos), whereas the runtime of \ac{QLBQM}, \ac{QFDA} and \ac{QFDAh} remains almost constant. The quality of the solutions is not significantly determined by the instance complexity. Further research is needed to investigate and quantify these observations in more detail.

Summarized, no general advantage of the quantum and quantum-inspired solvers was found. However, for certain instances the quantum-inspired hybrid usage of the Fujitsu digital annealer turned out to be a very promising alternative to Gurobi and was clearly superior to the usage of D-Wave’s hybrid Leap framework. Our study is not a conclusive result but rather an application-oriented case study that provides a snapshot of the current technology and leaves room for performance improvements on the modeling as well as the solver side. For example, an improvement of the quantum annealer inside the hybrid framework might be possible with additional problem-specific fine-tuning of the annealing schedule or other hardware-related parameters. Moreover, the recently released \ac{CQM} solver from D-Wave also promises to provide much better performance compared to the solver used in this work. Especially in an agile field such as quantum computing, a technology snapshot such as ours can hardly provide any forecasts about future developments. Therefore, in order to preserve an up-to-date assessment, further practical evaluations for real-world use cases will be necessary. The methods and results from this project can serve as a blueprint or at least point of reference for this kind of ongoing research.

\section{Acknowledgements}
\label{sec:acknowledgements}
We would like to thank Behrang Shafei, Jens Meissner and Horst Weiss for their invaluable input and support throughout the research process. Without their ongoing contributions, the work would not have been accomplished. This work was partly funded by the German Federal Ministry of Education and Research (Bundesministerium für Bildung und Forschung, BMBF) within the project ``Rymax One''.

\FloatBarrier
%\appendix
%\input{sections/appendix}

\FloatBarrier
\printbibliography

\end{document}